\documentclass[prb,twocolumn,floatfix,showpacs,amssymb]{revtex4}

\usepackage{graphicx}
\usepackage{dcolumn}
\usepackage{bm}
\begin{document}

\title{Thouless energy of a superconductor from non local
conductance fluctuations}
\author{S. Duhot}
\affiliation{
Centre de Recherches sur les Tr\`es Basses
Temp\'eratures (CRTBT),\\ CNRS, BP 166,
38042 Grenoble Cedex 9, France
}
\author{R. M\'elin}
\affiliation{
Centre de Recherches sur les Tr\`es Basses
Temp\'eratures (CRTBT),\\ CNRS, BP 166,
38042 Grenoble Cedex 9, France
}

\begin{abstract}
We show that
a spin-up electron from a normal metal
entering a superconductor propagates as
a composite object consisting of a spin-down hole 
and a pair in the condensate.
This leads to a factorization of the non local conductance
as two local Andreev reflections at both interfaces and one propagation
in the superconductor, which is tested numerically within a one dimensional
toy model of reflectionless tunneling.
Small area junctions are characterized by non local conductance
fluctuations. A treatment ignoring weak localization leads to
a Thouless energy inverse proportional to the sample size, as
observed in the numerical simulations. We show that weak localization
can have a strong effect, and leads to a coupling between
evanescent quasiparticles and the
condensate by Andreev reflections ``internal'' to the superconductor.
\end{abstract}

\pacs{74.50.+r,74.78.Na,74.78.Fk}

\maketitle

\section{Introduction}
Correlated pairs of electrons can be 
manipulated 
in solid state devices by extracting Andreev pairs from a
conventional superconductor,
being a condensate of Cooper pairs. This process is known as
Andreev reflection \cite{Andreev} at a normal metal / superconductor (NS)
interface.
One considers the future realization of devices designed for
manipulating separately one of the two electrons of an
Andreev pair and see the feedback on the other electron
\cite{Lambert,Jedema,Choi,Martin,Byers,Deutscher}.
The question arises of
exploring experimentally and
understanding theoretically the properties of the simplest
of these devices: a source of spatially
separated Andreev pairs propagating in different electrodes
forming, in short, ``non local'' Andreev pairs. 
The possibility of realizing
a source of non local Andreev pairs has indeed aroused a considerable
interest recently, both theoretical
\cite{Lambert,Jedema,Choi,Martin,Byers,Deutscher,Samuelson,Prada,Koltai,Melin-Peysson,Falci,Feinberg-des,Cht,EPJB,Melin-Feinberg-PRB,Tadei,Pistol,Melin-cond-mat}
and experimental \cite{Beckmann,Russo}.

In a theoretical prediction
prior to the experiments
\cite{Beckmann,Russo}, Falci {\it et al.} \cite{Falci} have obtained
from lowest order perturbation theory in the tunnel amplitudes
a vanishingly small non local signal with normal metals.
Russo {\it et al.} \cite{Russo}
have obtained on the contrary
a sizeable experimental non local signal
in a three terminal structure consisting of a
normal metal / insulator / superconductor / insulator / normal metal
(NISIN) trilayer.
The goal of our article is to provide a theory that,
together with Ref.~\onlinecite{Melin-cond-mat},
contributes to the understanding of
this experiment \cite{Russo}, as well as
related possible future experiments
on non local conductance fluctuations, and be consistent with
the other available 
experiment by Beckmann {\it et al.} with ferromagnets
\cite{Beckmann}. 

\begin{figure}
\begin{center}
\includegraphics [width=1. \linewidth]{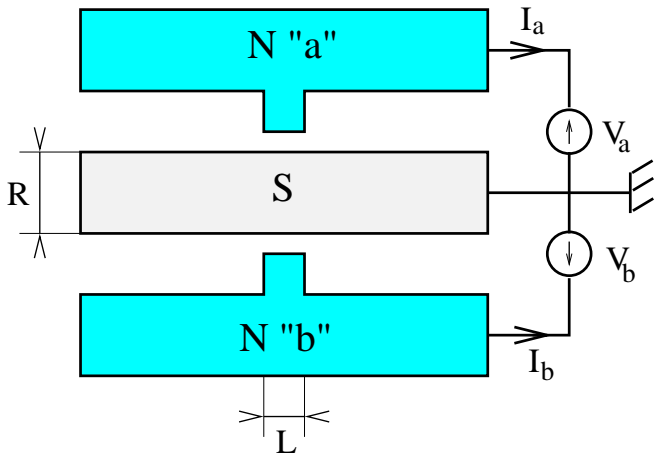}
\end{center}
\caption{(Color online.)
Schematic representation of the electrical circuit corresponding to 
the NISIN double interface interpolating between a localized contact
for $L\sim\lambda_F$ (with $\lambda_F$ the Fermi wave-length)
and and extended interface for $L\gg L_{\rm th}(\omega)$,
where $L_{\rm th}$ is the Thouless length corresponding to the
energy $e V_b$.
The current $I_a$ through electrode
``a'' is determined in response to a voltage $V_b$ on electrode ``b'',
with $V_a=0$. The dimensions $R$ (superconductor thickness) and $L$
(dimension of the junction) are shown on the figure.
The available experiment by Russo {\it et al.} \cite{Russo} corresponds to
extended interface with $L$ of the order of $0.5\mu$m and
$R\simeq 15 \div 200$nm.
\label{fig:schema-NISIN}
\label{fig:trilayer}
}
\end{figure}

Falci {\it et al.} \cite{Falci} have discussed the two competing
channels contributing to non local transport. An incoming electron
in electrode ``b'' can be transmitted as an electron in
electrode ``a'', corresponding to normal transmission
in the electron-electron channel (see the device on
Fig.~\ref{fig:trilayer} for the labels ``a'' and ``b'').
Conversely, it
can be transmitted as a hole in electrode ``a'' while a Cooper
pair is transfered in the superconductor. 
The later
transmission in the electron-hole channel corresponds
to a dominant ``non local''
Andreev reflection channel that can lead to 
spatially separated, spin entangled pairs of electrons.
The outgoing particles
in the two transmission channels have an opposite charge, resulting in
a different sign in the contribution to the current in electrode ``a''.
With normal metals, not only have the elastic cotunneling and
crossed Andreev reflection an opposite sign in the non local conductance,
but they are exactly opposite
within lowest order perturbation theory in the tunnel amplitudes.

It was already established that non local transport is dominated
by elastic cotunneling for localized interfaces \cite{Melin-Feinberg-PRB}.
The superconductor
can essentially be replaced by an insulator for a very
thin superconductor 
connected by tunnel contacts to a normal metal (assuming that the
superconductor can still be described by BCS theory).
We show that this picture breaks down if the
superconductor thickness is larger than the coherence length
because transport is mediated by
composite objects made of evanescent quasiparticles and pairs in the
condensate.

On the other hand, we find that small 
area junctions are controlled by a different physics 
with fluctuations of the non local conductance.
We find on the basis of an evaluation of the diffuson
in a superconductor, that the Thouless energy is inverse
proportional to the system size, which matches our
numerical simulations. We find also a possible large
coupling to the condensate provided by weak localization in the
superconductor.

The article is organized as follows.
The factorization of non local processes 
as two local Andreev reflections and a non local
propagation is discussed in  Sec.~\ref{sec:eh}.
The factorization of the non local conductance
is illustrated in Sec.~\ref{sec:1d}
in the case of one dimensional models (the Blonder, Thinkham,
Klapwijk (BTK) model\cite{BTK} and a Green's function model). 
The Thouless energy
of non local conductance fluctuations is examined in
Sec.~\ref{sec:anal-res} on the basis of the evaluation of
the diffuson. 
Numerical simulations are presented in Sec.~\ref{sec:num}.
The role of weak localization is pointed out in Sec.~\ref{sec:effective}.
Concluding remarks are given
in Sec.~\ref{sec:conclu}.

\section{Factorization of the non local resistance}
\label{sec:eh}
\label{sec:II}
\subsection{Existing results for $e V_b \gg E_{\rm th}(L)$}
\label{sec:II.A.}
\begin{figure}
\begin{center}
\includegraphics [width=1. \linewidth]{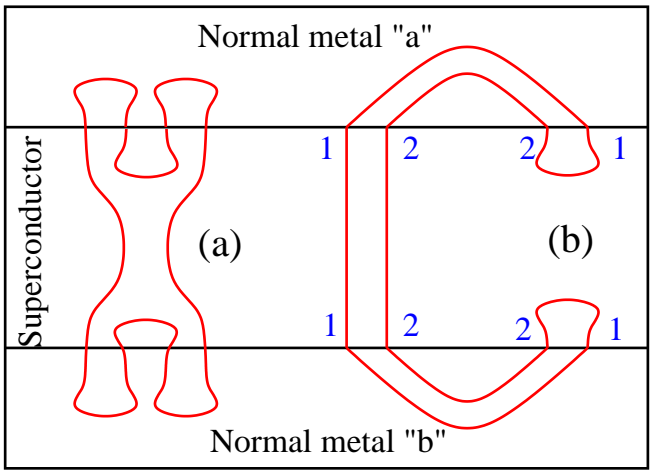}
\end{center}
\caption{(Color online.)
The diagrams representing the lowest order processes of order $T^4$.
The diagram (a), local with respect to a propagation in
the normal electrodes, was introduced in Ref.~\onlinecite{Melin-Feinberg-PRB}.
The diagram (b) is its non local counterpart. ``1'' and ``2'' 
correspond to the electron and hole
Nambu labels. The diagram on (b) factorizes in
two Andreev reflections at both interfaces, and a non local
propagation in the superconductor. The electron line crosses $8$
times the interfaces, so that the diagrams are of order $T^4$,
where $T\propto (t/\epsilon_F)^2$ is the normal transparency,
with $t$ the tunnel amplitude and $\epsilon_F$ the Fermi energy.
\label{fig:diagram}
}
\end{figure}

The diagram corresponding to the non vanishing lowest order process of order
$T^4$ (with $T$ the normal transparency)
is shown on Fig.~\ref{fig:diagram}a. 
This diagram 
is local with respect to excursions
parallel to the interfaces
if the bias voltage energy $e V_b$ is much larger than the
Thouless energy $E_{\rm th}(L)$ associated to the dimension $L$
of the junction parallel to the interface (see Fig.~\ref{fig:trilayer}),
as it is the case in the experiment
by Russo {\it et al.} \cite{Russo}. 
The corresponding non local conductance
\begin{equation}
\label{eq:Gab-def}
\overline{\cal G}_{a,b}(V_b)=\frac{\partial I_a}{\partial V_b}(V_b)
,
\end{equation}
where $I_a$ and $V_b$ are defined on Fig.~\ref{fig:trilayer},
is given by
\cite{Melin-Feinberg-PRB,Melin-cond-mat}
\begin{equation}
\label{eq:Gab-T4-final}
\overline{\cal G}_{a,b}(V_b)=-\frac{e^2}{h}
N_{\rm ch} T^4 \frac{\xi}{l_e^{(S)}}
\frac{\Delta^2}{\Delta^2-(e V_b)^2}
\exp{\left[-\left(\frac{2 R}{\xi}\right)\right]}
,
\end{equation}
where $N_{\rm ch}$ is the number of conduction channels,
$\Delta$ the superconducting gap, $\xi$ the superconducting
coherence length, $l_e^{(S)}$ the
superconductor elastic mean free path, $T$ the normal
local transparency, and the overline is an average over
disorder.
The local Andreev conductance is given by
\begin{equation}
\label{eq:G-loc}
{\cal G}_{\rm loc}(V_b)=2 \frac{e^2}{h} N_{\rm ch} T^2 
\frac{\Delta^2}{\Delta^2-(e V_b)^2}
,
\end{equation} 
where we used the ballistic result without disorder
because of the condition $e V_b \gg E_{\rm th}(L)$.

The resistance matrix probed in the experiment \cite{Russo}
is the inverse of the conductance matrix calculated theoretically:
\begin{equation}
\left[ \begin{array}{cc}
{\cal R}_{a,a}(V_b) & {\cal R}_{a,b}(V_b) \\
{\cal R}_{b,a}(V_b) & {\cal R}_{b,b}(V_b) \end{array}
\right]=
\left[ \begin{array}{cc}
\overline{\cal G}_{a,a}(V_b) & \overline{\cal G}_{a,b}(V_b) \\
\overline{\cal G}_{b,a}(V_b) & \overline{\cal G}_{b,b}(V_b) \end{array}
\right]^{-1}
,
\end{equation}
from what we deduce that the
non local resistance ${\cal R}_{a,b}(V_b)$ is  given by
\begin{equation}
{\cal R}_{a,b}(V_b) = \frac{-\overline{\cal G}_{a,b}(V_b)}
{ [{\cal G}_{\rm loc}(V_b)]^2-
\overline{\cal G}_{a,b}(V_b) \overline{\cal G}_{b,a}(V_b)}
,
\end{equation}
that simplifies into
\begin{equation}
\label{eq:Rab}
{\cal R}_{a,b}(V_b) \simeq -\frac{\overline{\cal G}_{a,b}(V_b)}
{\left[{\cal G}_{\rm loc}(V_b)\right]^2}
\end{equation}
if the thickness of the superconductor is larger than the
superconducting coherence length $\xi$, and leads to
\begin{eqnarray}
\nonumber
{\cal R}_{a,b}(V_b)
&=&
\frac{1}{4 N_{\rm ch}} \frac{h}{e^2} 
\left(\frac{\xi}{l_e^{(S)}}\right)\\
&\times&
\left(\frac{\Delta^2-(e V_b)^2}{\Delta^2}\right)
\exp{(- 2 R/\xi)}
.
\end{eqnarray}
The non local resistance at low bias is positive (dominated by elastic
cotunneling), as found in Ref.~\onlinecite{Melin-Feinberg-PRB}.
The case of extended interfaces is addressed 
in Ref.~\onlinecite{Melin-cond-mat}.

\begin{figure*}
\begin{center}
\includegraphics [width=1. \linewidth]{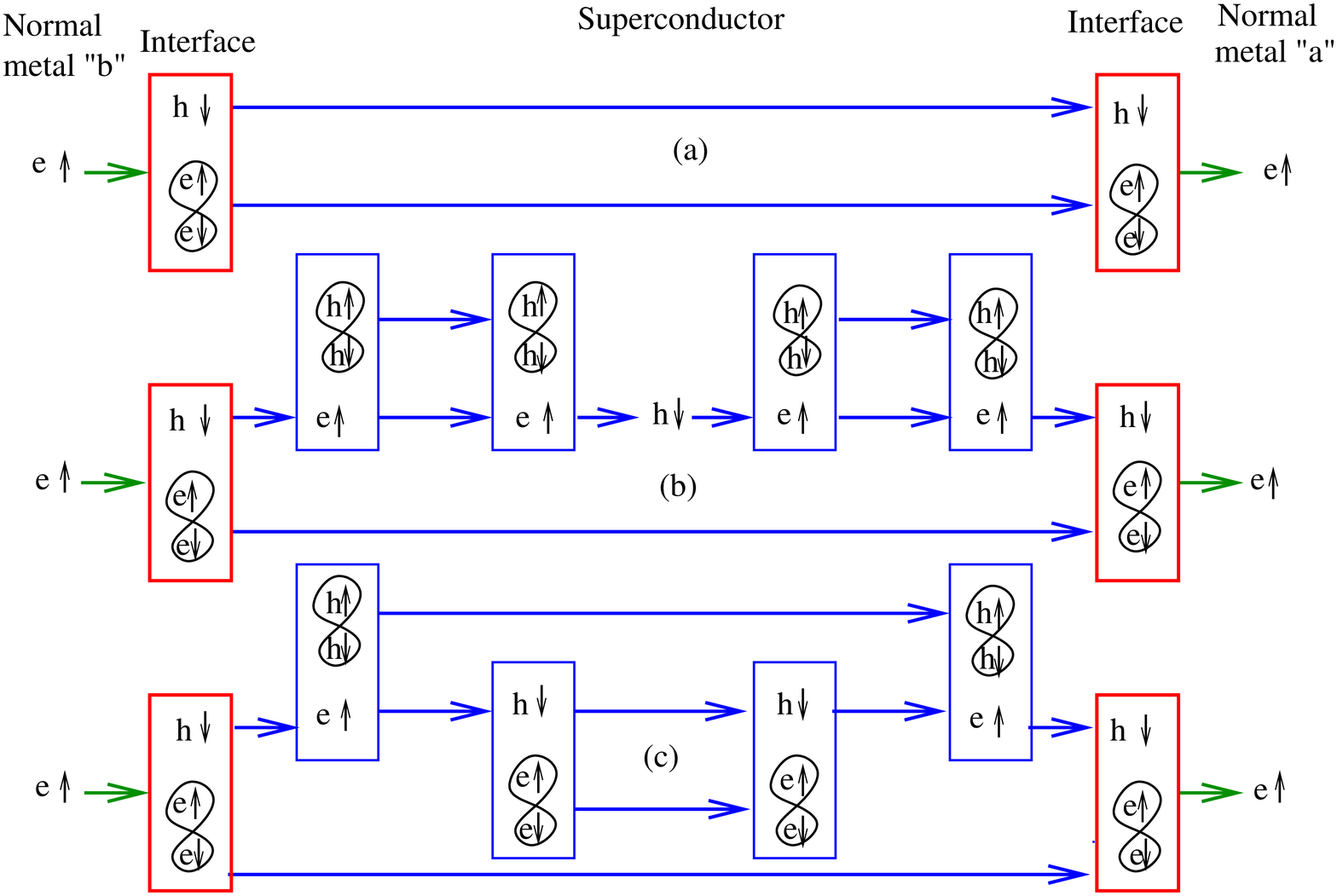}
\end{center}
\caption{(Color online.) Schematic representation of (a): the
formation of composite objects at the interfaces with
a ballistic superconductor;
(b): the sequential conversion of the composite objects
in the bulk of the superconductor in the absence of weak localization;
(c): the scattering induced by weak localization.
\label{fig:pedago}
}
\end{figure*}
\subsection{Case $e V_b \alt E_{\rm th}(L)$}
\label{sec:condition}
Now, if the bias voltage energy $e V_b$ is smaller than the
Thouless energy $E_{\rm th}(L)$, the
lowest order diagram of order $T^4$ becomes non local in the normal
electrodes (see Fig.~\ref{fig:diagram}b).
The diagram on Fig.~\ref{fig:diagram}b
corresponds to two Andreev reflections
in the normal electrodes, connected by a propagation in the superconductor,
so that the non local conductance factorizes into
\begin{equation}
\overline{\cal G}_{a,b}(V_b)=\frac{{\cal S}(V_b)
\left[{\cal G}_{\rm loc}(V_b)\right]^2}{ N_{\rm ch}}
,
\end{equation}
where ${\cal S}(V_b)$ is a transmission coefficient of the superconductor.
Using Eq.~(\ref{eq:Rab}), we find that
the crossed resistance
\begin{equation}
\label{eq:RS}
{\cal R}_{a,b}(V_b)= -\frac{{\cal S}(V_b)}{N_{\rm ch}}
\end{equation}
does not depend on the local conductances.
The scaling between
the local and non local conductances is tested in Sec.~\ref{sec:1d}
for the generalization of the model of reflectionless tunneling
at a single interface
introduced by Melsen and Beenakker \cite{Melsen}.

The factorization of the Andreev reflections at both interfaces suggests
that part of the current is carried by pairs in the condensate.
We thus arrive at
the notion of the transport of a composite object made
of an evanescent quasiparticle and a pair in the condensate:
an electron from a normal electrode
is transmitted in the superconductor as
a quasi-hole and a pair in the condensate (see
Fig.~\ref{fig:pedago}a). The consequences of this
qualitative picture are
considered below.

Finally, we note that the factorization of two Andreev reflections
at the interfaces is also valid
if $e V_b \gg E_{\rm th}(L)$ (see Sec.~\ref{sec:II.A.}),
as in the experiment by Russo {\it et al.}
\cite{Russo}. This is
because the normal Green's functions
are vanishingly small at zero energy in a superconductor.

\begin{figure}
\begin{center}
\includegraphics [width=1. \linewidth]{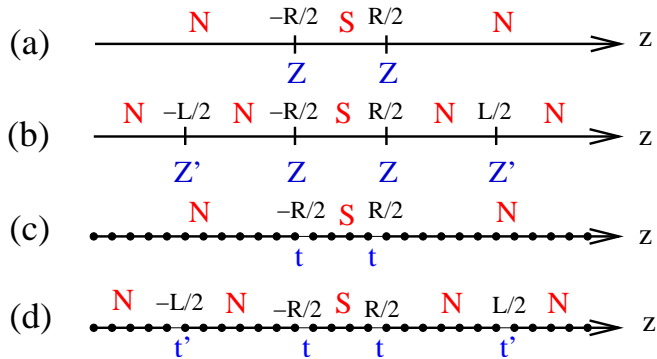}
\end{center}
\caption{(Color online.) Schematic representations of the
one dimensional models: (a) the BTK model for NISIN and
(b) NINISININ junctions, and (c) the tight-binding model for
NISIN and (d) NINISININ junctions.
\label{fig:schem-btk}
}
\end{figure}
\section{One dimensional models}
\label{sec:1d}
\subsection{Blonder, Tinkham, Klapwijk (BTK) approach}
\label{sec:btk}
\subsubsection{Non local conductance}
Let us consider now a one dimensional model of NISIN double
interface within the BTK approach \cite{japs} (see Fig.~\ref{fig:schem-btk}a).
The goal is two-fold: i) obtain the expression of the pair current
in the superconductor, and ii) test the factorization of the non
local conductance in the case of the model of reflectionless tunneling
introduced by Melsen and Beenakker \cite{Melsen}.

The gap of the superconductor is supposed to have a step-function
variation as a function of the coordinate $z$ along the chain:
$\Delta(z)=\Delta \theta(z+R/2)
\theta(R/2-z)$, and we suppose $\delta$-function scattering potentials
at the interfaces:
$V(z)=H \delta(z+R/2)+H \delta(z-R/2)$
\cite{BTK}. 
The interface transparencies are characterized by the
parameter $Z=2mH/\hbar^2 k_F$, where $v_F=\hbar k_F/m$ is the
Fermi velocity, with $m$ the electron mass and $k_F$ the Fermi
wave-vector. The one dimensional model is a simplified version
of the genuine three terminal geometry with a
supercurrent flow. 
The current in the normal
electrode ``a''
is not equal to the injected current in electrode ``b''
because part of the injected current has been converted
in a supercurrent.

The unknown coefficients in the expression of the wave-function
are determined from the
matching conditions at the interfaces
\cite{BTK}. Of particular interest are the amplitudes $a'(k_F R)$
and $b'(k_F R)$ of transmission in the electron-hole and electron-electron
channels from one normal metal to the other, corresponding
respectively to elastic cotunneling and non local Andreev reflection. 
Assuming $R\gg \xi$,
we expand $a'$ and $b'$ to lowest order in $\exp{(-R/\xi)}$, to find
the transmission coefficients
\begin{eqnarray}
&&\int_0^{2\pi} \frac{d (k_F R)}{2 \pi} |a'(k_F R)|^2 =\\
\nonumber
&& \left( \frac{1}{2 Z^4}-\frac{1}{2 Z^6}+\frac{1}{2 Z^8}
+ ...\right) e^{-2R/\xi} 
+ {\cal O}\left(e^{-4R/\xi}\right)\\
&&\int_0^{2\pi} \frac{d (k_F R)}{2 \pi} |b'(k_F R)|^2
= \\
\nonumber
&&\left( \frac{1}{2 Z^4}-\frac{1}{2 Z^6}+\frac{5}{4 Z^8}
+ ...\right) e^{-2R/\xi}
+ {\cal O}\left(e^{-4R/\xi}\right)
\end{eqnarray}
at $\omega=0$.
We deduce the first non vanishing term in the large-$R$,
large-$Z$ expansion of the non local transmission:
\begin{eqnarray}
\label{eq:Tprime}
T'&=&\int_0^{2\pi} \frac{d (k_F R)}{2 \pi} \left(|a'(k_F R)|^2
-|b'(k_F R)|^2\right)\\
\nonumber
&=&- \frac{3}{4 Z^8} e^{-2 R/\xi} +
{\cal O}\left(e^{-4R/\xi}\right)
.
\end{eqnarray}
In agreement with the Green's function approach
\cite{Melin-Feinberg-PRB,Melin-cond-mat}
corresponding to
the diagrams on Fig.~\ref{fig:diagram}, the non local 
conductance is dominated by elastic cotunneling and appears at
order $Z^{-8} \sim T^4$.
In agreement with Ref.~\onlinecite{Melin-Feinberg-PRB},
we find no non local Andreev reflection for
highly transparent interfaces corresponding to $Z=0$.

\begin{figure*}
\begin{center}
\includegraphics [width=.7 \linewidth]{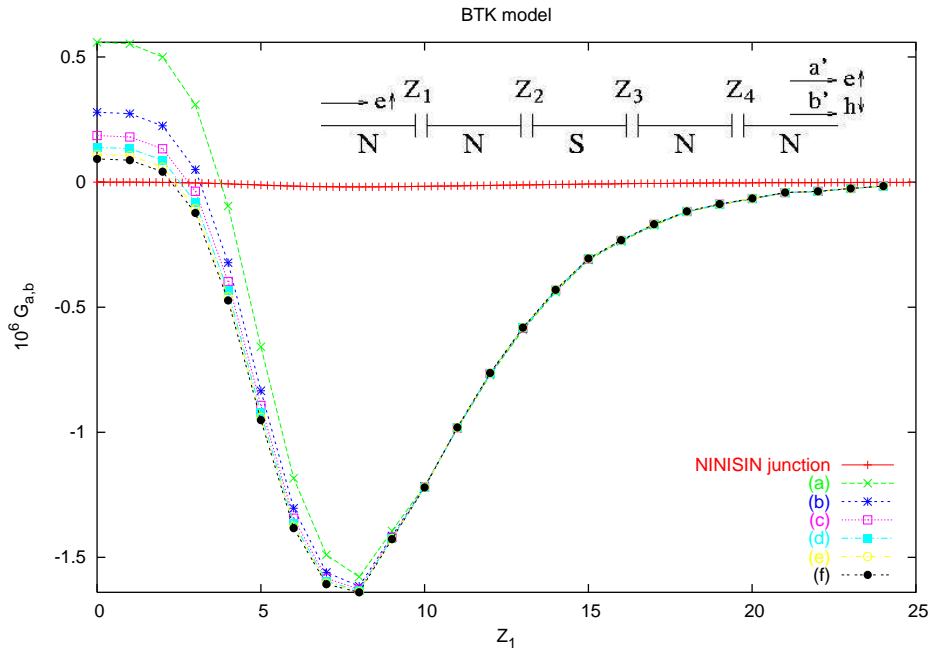}
\end{center}
\caption{(Color online.) Variation of the
non local conductance $G_{a,b}$ (in units of
$e^2/h$) for the junction on Fig.~\ref{fig:schem-btk}c, with
$Z'=Z_1=Z_4$ and $Z=Z_2=Z_3=10$. 
(a) ... (f) correspond to an increasing
values of the precision in the evaluation of the Fermi
phase factors related to the superconductor.  We have also
shown the much smaller
non local conductance of the NINISIN junction,
as a function of $Z_1$ for the NIN contact, with
the same value of $Z$ for the NIS contacts.
\label{fig:result1}
}
\end{figure*}

\subsection{Reflectionless tunneling}
\subsubsection{BTK approach}
\label{sec:NINISININ}
To discuss the form (\ref{eq:RS}) of the crossed resistance,
we include now multiple scattering in the normal electrodes and
consider two additional scatterers 
at positions $z_1=-L_1/2$ in the left electrode
and $z_2=L_2/2$ in the right electrode,
described by
the potentials $V'(z)=H' \delta(z-z_1)
+H' \delta(z-z_2)$, and leading to the
barrier
parameter $Z'=2mH'/\hbar^2 k_F$
(see Fig.~\ref{fig:schem-btk}b for the
definitions of $Z$ and $Z'$). This constitutes, for a double
interface, the analog of the model introduced by
Melsen and Beenakker \cite{Melsen} for a single interface.
We average numerically the non local
transmission coefficient over the Fermi oscillation phases
$\varphi_1=k_F(R-L_1)/2$, $\varphi=k_F R$ and
$\varphi_2=k_F(L_2-R)$. 

The variations of the non local conductance at zero bias
as a function of $Z'$ for a fixed $Z$
are shown on Fig.~\ref{fig:result1}, as well as the
corresponding non local conductance for the NINISIN junction.
The negative non local conductance
at small $Z_1$ 
for the NINISININ junction disappears when increasing the
precision of the integrals.
The variation of the non local conductance
on Fig.~\ref{fig:result1} shows a strong enhancement
by the additional scatterers, like reflectionless tunneling at
a single NIS interface \cite{Melsen}. 
\begin{figure}
\begin{center}
\includegraphics [width=1. \linewidth]{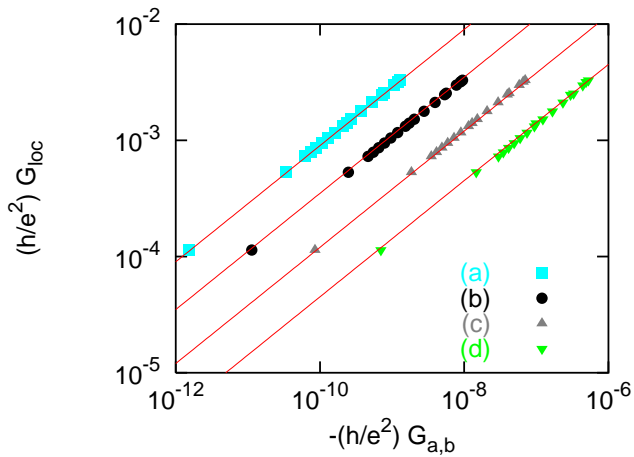}
\end{center}
\caption{(Color online.) Correlation between the nonlocal
conductance ($x$-axis)
and the local conductance ($y$-axis) for the peaks similar
to Fig.~\ref{fig:result1} as a function of $t'/T$ for a fixed
$t/T=0.05$ and
(a) $R/\xi=5$, (b) $R/\xi=4$, (c) $R/\xi=3$, and (d) $R/\xi=2$. 
The solid line is a fit to 
${\cal G}_{a,b}(t'/T) \propto \left[G_{\rm loc}(t'/T)\right]^2$.
\label{fig:scaling}
}
\end{figure}

\subsubsection{Green's functions: scaling between the local and
non local conductances}
\label{sec:green}

Considering the tight-binding model within Green's functions, the
variation of the non local
conductance of the NINISININ junction as a function of $t'$ for a fixed $t$
(see Fig.~\ref{fig:schem-btk}d) is similar to the BTK
model. Imposing the same normal conductance
in the BTK and in the tight-binding models leads to  $Z=(1-(t/T)^2)/(2 t/T)$,
where $T$ is the bulk hopping amplitude. The identification of $Z$
to $t/T$ results in a good (but not perfect)
agreement for the non local conductance
when the tight-binding and BTK results are rescaled on
each other.
The non local conductance ${\cal G}_{a,b}(V_b=0,t/T,t'/T)$
is shown on Fig.~\ref{fig:scaling} as a function of
the local conductance ${\cal G}_{\rm loc}(V_b=0,t/T,t'/T)$, fitted by
${\cal G}_{a,b}(V_b=0,t/T,t'/T)={\cal S}(V_b=0) 
\left[{\cal G}_{\rm loc}(V_b=0,t/T,t'/T)\right]^2$, corresponding
to Eq.~(\ref{eq:RS}) for $N_{\rm ch}=1$. The scaling
is very well obeyed, showing 
the validity of form (\ref{eq:RS}) of the crossed
resistance involving the destruction of a pair in the condensate
at one interface, its propagation in the superconductor
and its creation at the other interface.

\section{Thouless energy of a disordered superconductor}
\label{sec:anal-res}

\subsection{Relevance to experiments}
\label{sec:relevance}
We consider now non
local conductance fluctuations.
The total non local transmission coefficient is given by
$T_{tot}=T_{e-e}-T_{e-h}$, where $T_{e-e}$ and
$T_{e-h}$ are the transmission coefficients in the
electron-electron and electron-hole channels respectively.
As discussed in Sec.~\ref{sec:II}, one has
$\overline{T_{tot}}=0$ but
\begin{equation}
\overline{\left(T_{tot}\right)^2}
=\overline{\left(T_{e-e}\right)^2}
+ \overline{\left(T_{e-h}\right)^2}
-2 \overline{\left(T_{e-e}T_{e-h}\right)^2}
.
\end{equation}
Inspecting the corresponding lowest order diagrams shows that
$\overline{T_{e-e} T_{e-h}}=-\overline{\left(T_{e-e}\right)^2}$, where
we suppose that the normal metal phase coherence length
is vanishingly small, therefore avoiding the specific
effects of extended interfaces \cite{Melin-cond-mat}.
The root mean square of the non local conductance fluctuations is thus
proportional to $(e^2/h)T^2$ while the average non local conductance is 
proportional to $(e^2/h)T^4 N_{\rm ch}$ (see Eq.~\ref{eq:Gab-T4-final}).
The fluctuations are important
for small junctions such that $T^2 N_{\rm ch}\alt 1$. 
\begin{figure}
\includegraphics [width=1 \linewidth]{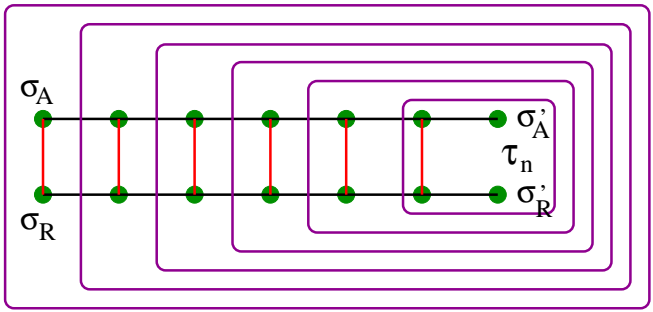}
\caption{(Color online.) Schematic representation of the recursive
calculation of the matrix diffuson in the ladder approximation
used by Smith and Ambegaokar 
\cite{Smith}. The integrals over the wave-vectors are carried out
recursively, starting from the right of the diagram. The two horizontal
black lines correspond to the advanced and retarded Green's functions.
The green dots correspond to the disorder vertices, and the
vertical red lines correspond to the impurity lines. We
have shown the Nambu labels $(\sigma_A,\sigma_R)$
and $(\sigma'_A,\sigma'_R)$ at the extremities of the diffuson.
\label{fig:diffuson}
}
\end{figure}

\subsection{Diffusons in a superconductor}
\label{sec:diff}
\subsubsection{Evaluation of the diffusons}
\label{sec:VA1}
Let us first evaluate the Thouless energy of a superconductor in
the absence of crossings between diffusons.
Smith and Ambegaokar \cite{Smith}
start from one extremity of the ladder diagram
and calculate
recursively the integrals over the wave-vectors. Once the right-most
integral on Fig.~\ref{fig:diffuson} has been evaluated, one is
left with a ``ladder'' with one less rung, but with a different
$2\times 2$ matrix at the extremity. The four parameter recursion
relations reduce to a matrix geometric series
in the sector $(\hat{\tau}_0,\hat{\tau}_1)$, and to
another matrix geometric series in the sector
($\hat{\tau}_2$,$\hat{\tau}_3$), where
$\hat{\tau}_n$ are the four Pauli matrices, with $\hat{\tau}_0$
the identity.
 
More precisely, we define the four matrix diffusons
\begin{equation}
\label{eq:D}
\overline{\hat{\cal D}}_{q,\delta \omega}(\hat{\tau}_n)=
v^2 \int \frac{d^3{\bf k}}{(2\pi)^3}
\overline{\hat{\tau}_3 \hat{G}({\bf k},\omega) \hat{\tau}_n
\hat{G}({\bf k}+{\bf q},\omega+\delta \omega)\hat{\tau}_3}
,
\end{equation}
where $n=0,...,3$, $q=|{\bf q}|$ is the modulus of the wave-vector,
and $\delta \omega$ 
is small compared to the energy $\omega$.
The microscopic disorder scattering potential is given by
$v^2=4\pi \epsilon_F/\tau_e$, with $\epsilon_F$ the Fermi energy
and $\tau_e$ the elastic scattering
time, related to the elastic scattering length $l_e$ by the relation
$l_e=v_F \tau_e$.
We find
\begin{eqnarray}
\label{eq:diff1}
\overline{\hat{\cal D}}_{q,\delta \omega}(\hat{\tau}_0)&=&
X \left(\Delta^2 \hat{\tau}_0-\omega \Delta \hat{\tau}_1\right)\\
\overline{\hat{\cal D}}_{q,\delta \omega}(\hat{\tau}_1)&=&
X \left(\omega \Delta \hat{\tau}_0-\omega^2 \hat{\tau}_1\right)
,
\end{eqnarray}
in the sector $(\hat{\tau}_0,\hat{\tau}_1)$, and
\begin{eqnarray}
&&\overline{\hat{\cal D}}_{q,\delta \omega}(-i\hat{\tau}_2)=X \left(-
\frac{3 {\cal D}_0 \Delta \delta \omega \sqrt{\Delta^2-\omega^2}}
{2 v_F^2} \hat{\tau}_3- \right.\\
&&\left.i(\Delta^2-\omega^2)\hat{\tau}_2 \right)
\nonumber
\\
\label{eq:diff4}
&&\overline{\hat{\cal D}}_{q,\delta \omega}
(\hat{\tau}_3)= X {\cal D}_0 \delta \omega
\left(
-\frac{ {\cal D}_0^2 \Delta \delta \omega}{4 v_F^4}
\hat{\tau}_3-\frac{3 i \Delta}{2 v_F^2} \sqrt{\Delta^2-\omega^2} 
\hat{\tau}_2 \right)
,
\nonumber
\end{eqnarray}
in the sector $(\hat{\tau}_2,\hat{\tau}_3)$.
We used the notation
\begin{equation}
\label{eq:X}
\frac{1}{X}=
\frac{3 {\cal D}_0 (\Delta^2-\omega^2)}{v_F^2}
\left[\sqrt{\Delta^2-\omega^2}+\frac{{\cal D}_0 q^2}{4}-
\frac{\omega \delta \omega}{2\sqrt{\Delta^2-\omega^2}}\right]
,
\end{equation}
where ${\cal D}_0$ is the diffusion constant.

\subsubsection{Non local transmission coefficient}
The relation between the diffusons in the superconductor and
non local transport is provided by the
non local conductance (\ref{eq:Gab-def}).
The non local conductance ${\cal G}_{a,b}^{(2)}(\omega)$
of order $T^2$
is related to the transmission coefficients according to
\begin{equation}
\label{eq:Gab2}
{\cal G}_{a,b}^{(2)}(\omega)=
\frac{e^2}{h}
\left[T_{(1,1)}^{(1,1)}(\omega)-T_{(1,1)}^{(2,2)}(\omega)\right]
,
\end{equation}
with
\begin{eqnarray}
&&T_{(\sigma_A,\sigma_R)}^{(\sigma'_A,\sigma'_R)}(\omega)= T^2
\epsilon_F^2
\int \frac{d^3 {\bf q}}{(2\pi)^3} e^{i{\bf q}.{\bf R}}\\
&&\times \int \frac{d^3 {\bf k}}{(2\pi)^3} 
\overline{
\hat{G}^{\sigma_A,\sigma'_A}({\bf k},\omega)
\hat{G}^{\sigma'_R,\sigma_R}
({\bf k}+{\bf q},\omega)}
.
\nonumber
\end{eqnarray}
The notation $T_{(\sigma_A,\sigma_R)}^{(\sigma'_A,\sigma'_R)}(\omega)$
corresponds to the transmission coefficient related to a diffuson
with the Nambu labels $(\sigma_A,\sigma_R)$ for the advanced and retarded
propagators at one extremity, and the Nambu labels
$(\sigma'_A,\sigma'_R)$ at the other extremity
(see Fig.~\ref{fig:diffuson}).
The transmission coefficients
$\overline{T}_{(1,1)}^{(1,1)}(\omega)$ and
$\overline{T}_{(1,1)}^{(2,2)}(\omega)$ encode
elastic cotunneling and non local Andreev reflection respectively.
With the notations in Sec.~\ref{sec:relevance}, we have
$\overline{T}_{(1,1)}^{(1,1)}(\omega)=T_{e-e}(\omega)$
for transmission in the electron-electron channel, and
$\overline{T}_{(1,1)}^{(2,2)}(\omega)=T_{e-h}(\omega)$
for transmission in the electron-hole channel.
We deduce from Eq.~(\ref{eq:diff4}) that
$\overline{\cal G}^{(2)}_{a,b}(\omega)=0$:
the average non local conductance vanishes
to order $T^2$, in agreement both with Sec.~\ref{sec:II}
and with an early work \cite{Feinberg-des}
in the disordered case. The transmission coefficients
$\overline{T}_{(1,2)}^{(1,2)}(\omega)$ and
$\overline{T}_{(2,1)}^{(2,1)}(\omega)$ involve the propagation of
a pair in the condensate in parallel to the quasiparticle channels,
as in the diagram on Fig.~\ref{fig:diagram}b.

\subsection{Thouless energy}
The Thouless energy 
is defined from the non local conductance fluctuations
by the decay of the autocorrelation of the non local conductance
\begin{equation}
\label{eq:correT}
\langle \left[
\overline{{\cal G}_{a,b}(\omega) {\cal G}_{a,b}(\omega+\delta \omega)}
-\overline{\cal G}_{a,b}(\omega)
\overline{\cal G}_{a,b}(\omega+\delta \omega) \right]
\rangle_\omega
\end{equation}
as $\delta \omega$ increases, where 
$\langle ... \rangle_\omega$ 
denotes an average over the energy $\omega$ in a given window. 

The autocorrelation of
the non local conductance defined by Eq.~(\ref{eq:correT})
is related to the autocorrelation of the transmission
coefficients
\begin{eqnarray}
\label{eq:correTbis}
&&
\langle
\overline{T_{(\sigma_1,\sigma_2)}^{(\sigma'_1,\sigma'_2)}(\omega)
T_{(\sigma_3,\sigma_4)}^{(\sigma'_3,\sigma'_4)}(\omega+\delta \omega)}\\
&&-\overline{T}_{(\sigma_1,\sigma_2)}^{(\sigma'_1,\sigma'_2)}(\omega)
\overline{
T}_{(\sigma_3,\sigma_4)}^{(\sigma'_3,\sigma'_4)}(\omega+\delta \omega)
\rangle_\omega
.
\nonumber
\end{eqnarray}

More precisely, the non local conductance to lowest order in the
tunnel amplitudes is given by
\begin{equation}
{\cal G}_{a,b}(\omega)={\cal A}
\left[ g_{a,b}^{1,1,A} g_{b,a}^{1,1,R}
-g_{a,b}^{1,2,A}g_{b,a}^{2,1,R} \right]
,
\end{equation}
where ``1'' and ``2'' refer to the electron and hole
Nambu components, ``A'' and ``R'' stand for
advanced and retarded, $g_{a,b}^{1,1}$ is a propagation from ``a'' to ``b''
in the electron-electron channel, and $g_{a,b}^{1,2}$
in the electron-hole channel. The prefactor ${\cal A}$,
not directly relevant to our discussion, can be
found in Ref.~\onlinecite{Melin-Feinberg-PRB}. We find easily
\begin{eqnarray}
&&\overline{{\cal G}_{a,b}(\omega) {\cal G}_{a,b}(\omega+\delta \omega)}
={\cal A}^2 \sum_{k_1,...,k_4} e^{i(k_1-k_2+k_3-k_4)R} \\
\nonumber
&&\left[ \overline{g^A_{1,1}(k_1,\omega)
g^R_{1,1}(k_2,\omega) g^A_{1,1}(k_3,\omega+\delta\omega)
g^R_{1,1}(k_4,\omega+\delta \omega)}\right.\\
\nonumber
&-&\overline{g^A_{1,1}(k_1,\omega)
g^R_{1,1}(k_2,\omega) g^A_{1,2}(k_3,\omega+\delta\omega)
g^R_{2,1}(k_4,\omega+\delta \omega)}\\
\nonumber
&-& \overline{g^A_{1,2}(k_1,\omega)
g^R_{2,1}(k_2,\omega) g^A_{1,1}(k_3,\omega+\delta\omega)
g^R_{1,1}(k_4,\omega+\delta \omega)}\\
\nonumber
&+& \left. \overline{g^A_{1,2}(k_1,\omega)
g^R_{2,1}(k_2,\omega) g^A_{1,2}(k_3,\omega+\delta\omega)
g^R_{2,1}(k_4,\omega+\delta \omega)}
\right]
.
\end{eqnarray}
The quantity
\begin{equation}
{\cal S}_{a,b}(\omega,\omega+\delta\omega)=
\overline{ {\cal G}_{a,b}(\omega)
{\cal G}_{a,b}(\omega+\delta\omega)}
-\overline{ {\cal G}_{a,b}(\omega) {\cal G}_{a,b}(\omega)}
\end{equation}
is evaluated by discarding the Nambu components of the type
$\overline{g^A_{1,1}(k_1,\omega) g^A_{1,2}(k_2,\omega+\delta \omega)}$,
much smaller than $\overline{g^A_{1,1}(k_1,\omega) g^A_{1,1}
(k_2,\omega+\delta \omega)}$ and
$\overline{g^A_{1,2}(k_1,\omega) g^A_{2,1}
(k_2,\omega+\delta \omega)}$ if $\omega$ is small compared to $\Delta$.
In addition, we use
\begin{equation}
\overline{ {\cal G}_{a,b}(\omega) {\cal G}_{a,b}(\omega) }
\simeq
\overline{ {\cal G}_{a,b}(\omega+\delta\omega)
{\cal G}_{a,b}(\omega+\delta \omega) }
\end{equation}
within the small energy window that we consider.
We obtain
\begin{eqnarray}
&&{\cal S}_{a,b}(\omega,\omega+\delta\omega)=\\
\nonumber
&&2\left(T_{(1,1)}^{(1,1)}(R,\omega,\delta \omega)\right)^2
+ 2\left(T_{(1,1)}^{(2,2)}(R,\omega,\delta \omega)\right)^2\\
\nonumber
&-& 2\left(T_{(1,1)}^{(1,1)}(R,\omega,0)\right)^2
- 2\left(T_{(1,1)}^{(2,2)}(R,\omega,0)\right)^2
,
\end{eqnarray}
where the Fourier transforms with respect to the spatial variable
of $T_{(1,1)}^{(1,1)}(R,\omega,\delta \omega)$ and
$T_{(1,1)}^{(2,2)}(R,\omega,\delta \omega)$ are given by
\begin{eqnarray}
\label{eq:(x)}
&&T_{(1,1)}^{(1,1)}(q,\omega,\delta \omega)=
T_{(1,1)}^{(2,2)}(q,\omega,\delta \omega)=\\
&&\frac{v_F^2 \Delta^2}
{3 {\cal D}_0^2 (\Delta^2-\omega^2)
\left[\sqrt{\Delta^2-\omega^2}+{\cal D}_0q^2/4-\omega \delta \omega
/2\sqrt{\Delta^2-\omega^2}\right]}
\nonumber
,
\end{eqnarray}
deduced from Sec.~\ref{sec:VA1}.
The notation $R$ stands for the distance between the contacts
``a'' and ``b''.
Taking the Fourier transform of Eq.~(\ref{eq:(x)}), we 
obtain
\begin{eqnarray}
&&{\cal S}_{a,b}(\omega,\delta \omega)
\left[ \overline{{\cal G}_{a,b}(\omega)
{\cal G}_{a,b}(\omega)}\right]^{-1}\\
\nonumber
&=&\exp{\left(i \sqrt{3} \frac{R}{\xi}
\frac{\omega \delta \omega}{4(\Delta^2-\omega^2)}
\right)}-1
,
\end{eqnarray}
that dephases above the Thouless energy
\begin{equation}
E_c=\delta \omega_c= \frac{8\pi}{\sqrt{3}}
\frac{\Delta^2-\omega^2}{\omega} \frac{\xi}{R}
,
\end{equation}
for energies $\omega$ large compared to $\Delta
\sqrt{\xi/R}$, so that $E_c$ is much smaller than $\omega$.

\section{Numerical results}
\label{sec:num}

\subsection{The different length scales in the simulations}
The non local transport simulations are carried out in a quasi-1D
geometry, on a strip of longitudinal dimension $L$
and of transverse dimension $L_y$, corresponding to $M$
transverse modes.
We calculate non local transport along the $z$ direction.
The trilayer geometry with an aspect ratio
similar to the experiment by Russo {\it et al.} \cite{Russo},
would require
much larger system sizes to
have a reasonable separation between the different length
scales in the $y$ direction while $L$ is much larger than $L_y$.
The relevant length scales in the diffusive regime are
given by increasing order by the Fermi wave-length $\lambda_F$,
the elastic mean free path $l_e$, the superconducting
coherence length $\xi$ and the sample size.

\subsection{Ballistic system and small disorder}
We use typically $M=10$, $L/a_0$ ranging from $80$ to $100$
for a method \cite{Levy-Yeyati} based on
the inversion of the Dyson matrix, and much higher values of
$L/a_0$ (in units of the tight binding model lattice spacing $a_0$),
for a complementary method consisting in connecting together
several conductors by a hopping self-energy, given
the Green's functions of each conductor evaluated by
the inversion of the Dyson matrix.  Disorder is introduced
as in the Anderson model by a random potential between
$-W$ and $W$.

\begin{figure}
\includegraphics [width=1. \linewidth]{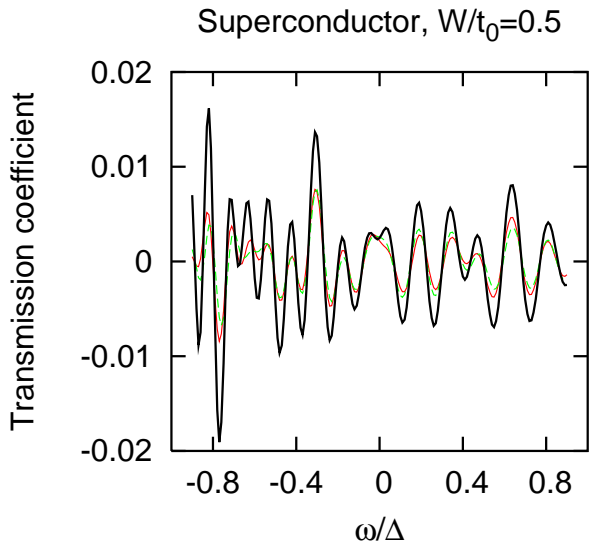}
\caption{(Color online.) Energy dependence
of the superconducting transmission coefficient $T'(\omega/\Delta)$
[defined by Eq.~(\ref{eq:Tp})]
through a diffusive superconductor on a strip with $L_y/a_0=10$
and $L/a_0=100$, in the limit of small disorder. 
The bold line corresponds to the ballistic result, and
the two other traces correspond to two realizations of disorder
with $W/t_0=0.5$ and $l_e/a_0\simeq 500$. The ballistic coherence
length is $\xi/a_0\simeq 33$. This figure has been 
obtained with a method based on the inversion of the Dyson matrix.
\label{fig:Fig3a}
}
\end{figure}

\begin{figure}
\includegraphics [width=1. \linewidth]{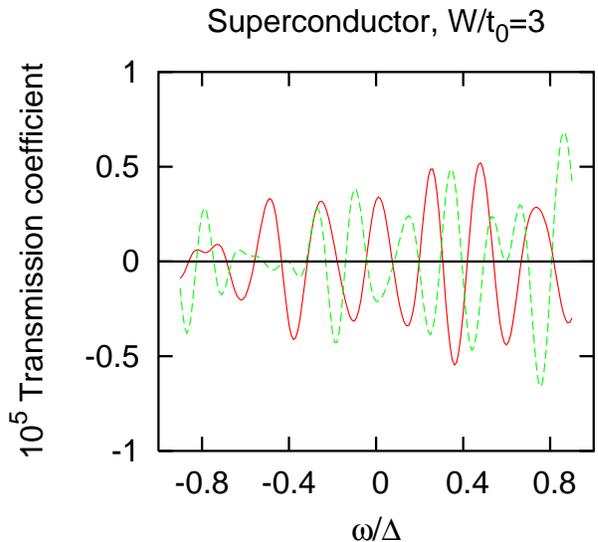}
\caption{(Color online.) Energy dependence
of the superconducting transmission coefficient $T'(\omega/\Delta)$
[defined by Eq.~(\ref{eq:Tp})]
through a diffusive superconductor on a strip with $L_y/a_0=10$
and $L/a_0=100$, in the diffusive limit. 
The two traces correspond to different realizations of
disorder corresponding to $W/t_0=3$ and $l_e/a_0\simeq 16$.
This figure has been 
obtained with a method based on the inversion of the Dyson matrix.
\label{fig:Fig3d}
}
\end{figure}


The normalized transmission coefficient $T'(\omega)$
that we calculate numerically is related to the non local conductance
by the relation
\begin{equation}
\label{eq:Tp}
T'(\omega)=
\frac{h}{e^2} T^{-2} {\cal G}_{a,b}^{(2)}(\omega)
,
\end{equation}
where ${\cal G}_{a,b}^{(2)}(\omega)$ is
the contribution of order $T^2$ to the non local conductance, with
$T$ the normal transparency.
The transmission coefficient $T'(\omega)$ defined by Eq.~(\ref{eq:Tp})
fluctuates around zero as a function of energy because the wave-vectors
of the different channels
vary with energy. The characteristic energy
scale in the oscillations of the transmission coefficient
is the ballistic normal state
Thouless energy associated to the
dimension $L$ (see Fig.~\ref{fig:Fig3a} in the forthcoming
section~\ref{sec:totoA}).

\begin{figure}
\includegraphics [width=1. \linewidth]{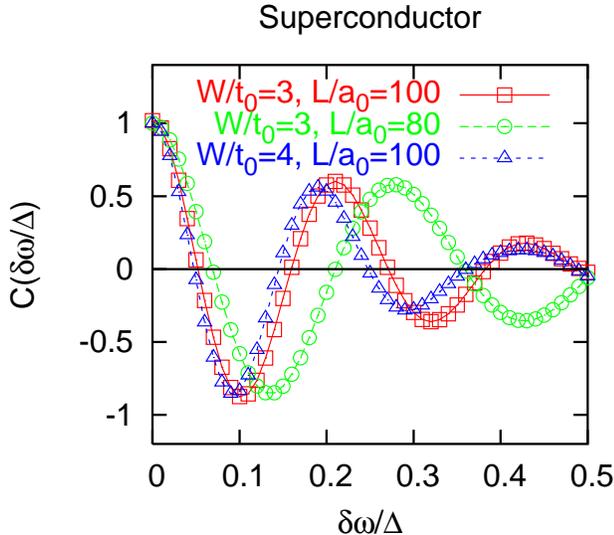}
\caption{(Color online.) 
Autocorrelation function of the transmission coefficient
[see Eqs.~(\ref{eq:auto1})-(\ref{eq:auto2})],
for ($W/t_0=3$, $L/a_0=80$), ($W/t_0=3$, $L/a_0=100$),
and ($W/t_0=4$,$L/a_0=100$). $W/t_0=3$ corresponds to
$l_e/a_0\simeq 16$, and $W/t_0=4$ corresponds to
$l_e/a_0\simeq 10$.
The errorbars are smaller than the size of the symbols.
\label{fig:autoco}
}
\end{figure}

\subsection{Thouless energy of a disordered superconductor}
\label{sec:totoA}

Figs.~\ref{fig:Fig3a} and \ref{fig:Fig3d}
show the energy dependence of the
superconducting transmission coefficient $T'(\omega/\Delta)$
defined by Eq.~(\ref{eq:Tp}).
The fluctuations of the transmission coefficient are close
to the ballistic limit result in the limit of small disorder
(see Fig.~\ref{fig:Fig3a}). We obtain regular
fluctuations of the transmission coefficient of
a superconductor in the diffusive limit where
the normal transmission coefficient is characterized
by fluctuations (see Fig.~\ref{fig:Fig3d}).
We used a large number of realizations of disorder at a single energy
to show
that the transmission coefficient averages to zero
because of disorder. This shows that the regular fluctuations in
the disordered system are genuinely related to disorder, and do not have the
same origin as in the ballistic system.

\begin{figure}
\includegraphics [width=1. \linewidth]{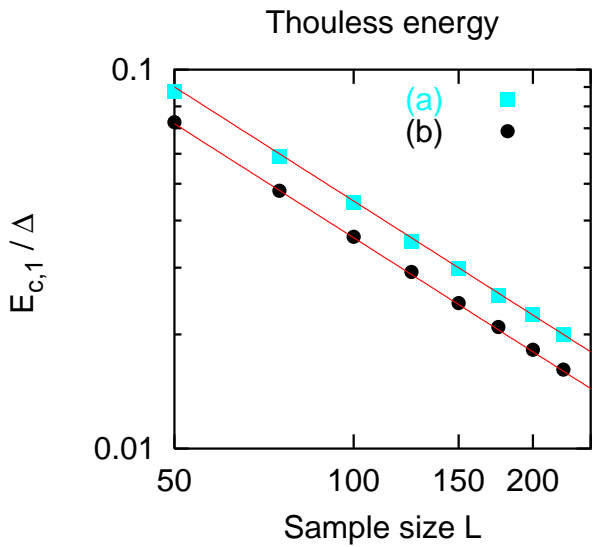}
\caption{(Color online.) 
Variation of the Thouless energy $E_{c,1}/\Delta$ as a function
of the system size $L/a_0$
in a log-log plot, for (a):
$W/t=1$ and $l_e/a_0\simeq 127$ (b): $W/t=2$ and $l_e/a_0
\simeq 34$. The solid lines are a fit to
$E_{c,a}/\Delta \sim 1/L$.
\label{fig:Ec}
}
\end{figure}

To characterize the regular fluctuations, we calculate
the normalized autocorrelation
of the transmission coefficient ${\cal C}(\delta \omega)=
\langle A(\omega,\delta \omega)/B(\omega,\delta \omega)\rangle_\omega$,
with
\begin{eqnarray}
\label{eq:auto1}
A(\omega,\delta \omega) &=&
\overline{T'(\omega+\delta \omega) T'(\omega)}\\
\label{eq:auto2}
B(\omega,\delta \omega) &=&
\sqrt{\overline{T'}(\omega+\delta \omega) \overline{T'}(\omega)}
,
\end{eqnarray}
where $\langle ...\rangle_\omega$ is an average over the energy
$\omega$,
as in Eq.~(\ref{eq:correT}).
The autocorrelation ${\cal C}(\delta \omega)$ is characterized by
oscillatory damped oscillations (see Fig.~\ref{fig:autoco}), in contrast
to the autocorrelation of conductance fluctuations in the
normal case that is damped without oscillations.
The energy scales $E_{c,1}$ and $E_{c,2}$
related to period of oscillations and to the damping 
increase as the system size decreases (see Fig.~\ref{fig:autoco}),
in agreement with the expected behavior for Thouless-like energies.
Going to larger system sizes, we find that
$E_{c,1}$ scales like the inverse of the
sample size (see Fig.~\ref{fig:Ec}).

The comparison between
Fig.~\ref{fig:Fig3d} for ($W/t_0=3$, $l_e/a_0\simeq 16$),
and similar data for ($W/t_0=2$,
$l_e/a_0\simeq 34$) and
($W/t_0=4$, $l_e/a_0\simeq 10$)
show that $E_{c,1}$ and $E_{c,2}$
have a weaker dependence on the elastic mean free
path than for a normal diffusive system.

\section{Effective scattering induced by weak localization}
\label{sec:effective}
\begin{figure*}
\includegraphics [width=1. \linewidth]{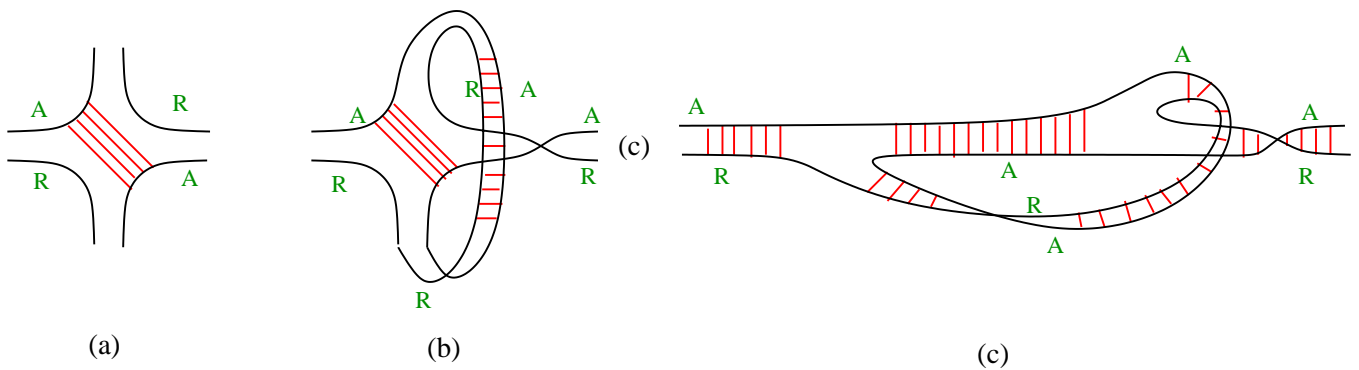}
\caption{(Color online.) (a) One of the
Gorkov-Larkin-Khmelnitskii-Hikami (GLKH) boxes \cite{Gorkov,Hikami,Montambaux}
in the superconducting case,
dressed by a diffuson. (b) The self-crossing of a diffuson with the
superconducting GLKH box. (c) Another representation of (b), with
two additional diffusons at the entry and exit of the GLKH box.
``A'' and ``R'' stand for advanced and retarded.
\label{fig:Hikami}
}
\end{figure*}

We find a formal analogy between the calculation of the non local
conductance fluctuations in the preceding section, and
the linear response theory of collective modes
\cite{Anderson,Bogoliubov,Kulik}. Namely,
the non local conductance fluctuations can be viewed as a generalized
susceptibility in linear response, but otherwise
the two models involve
rather different physical quantities. We show now that
weak localization can induce additional couplings to the condensate.
Evaluating all the Nambu labels at the two three-diffuson vertices
(see Fig.~\ref{fig:Hikami}c) is
possible in the limit $\omega \ll \Delta$ because of the constraint that
the normal local Green's functions can be discarded in this limit
that corresponds ``internal'' Andreev reflection
processes as on Fig.~\ref{fig:pedago}c.
The diagrams on Fig.~\ref{fig:Hikami}c then defines a set of 
transmission coefficients modified by weak localization.
With the notations
$A=T_{(1,2)}^{(1,2)}$, $B=T_{(1,2)}^{(2,1)}$,
$C=T_{(1,1)}^{(2,2)}$, $D=T_{(1,1)}^{(1,1)}$, we find for
the ``renormalized'' transmission coefficients
\begin{eqnarray}
\tilde{A}&=&A+\lambda \left[ A B (A+B)^2+D^2(A^2+B^2)\right.\\
\nonumber
&+&\left.2 ABC^2\right]+{\cal O}(\lambda^2)\\
\tilde{B}&=&B+\lambda \left[ A B (A+B)^2+C^2(A^2+B^2)\right.\\
\nonumber
&+&\left.2 ABD^2\right]+{\cal O}(\lambda^2)\\
\tilde{C}&=&C+\lambda D(A+B)(D^2+3 C^2)
+{\cal O}(\lambda^2)\\
\tilde{D}&=&D+\lambda C(A+B)(C^2+3 D^2)
+{\cal O}(\lambda^2)
,
\end{eqnarray}
with the perturbative parameter
$\lambda\sim (\tau^2 \epsilon_F \Delta)^{-4}$ that can turn out
to be large. Weak localization can thus lead
to a large effective scattering for the processes on
Fig.~\ref{fig:pedago}c with multiple imbricate Andreev reflections
providing a coupling between the condensate and
the evanescent quasiparticle channels.

\section{Conclusions}
\label{sec:conclu}

To conclude, we have provided a theory of non local 
conductance fluctuations
at normal metal / superconductor double interfaces. 
First, reconsidering the case of the average non local conductance,
we confirm that the central
role is played by higher order processes in the tunnel amplitude.
We found that for these processes
part of the non local current circulates as
pairs in the condensate, not only as evanescent quasiparticles.
The crossed conductance at zero bias factorizes in the Andreev
conductances at the two interfaces, and a factor related to the
propagation in the superconductor. This factorization was
tested in the context of a model 
of reflectionless tunneling \cite{Melsen}.

On the other hand we found numerically
regular fluctuations of the non local conductance.
The Thouless energy inverse proportional to the system size
obtained in the simulations can be
interpreted by a model ignoring weak localization. Alternatively,
an energy scale inverse proportional to the system size could
have received an interpretation in terms of the
Anderson\cite{Anderson}-Bogoliubov\cite{Bogoliubov}
collective mode, which is not in contradiction with the fact that
weak localization can induce a strong coupling
to the condensate if the superconductor 
elastic mean free path is sufficiently small. However, 
disorder in our simulations is most likely not strong enough to 
correspond to this possibility.
Finally, the Thouless energy of a normal cavity appears also in circuit
theory\cite{Morten}.
Our model for the conductance fluctuations clarifies the
concept of Thouless energy intrinsic to a superconductor.

\section*{Acknowledgements}

We thank D. Feinberg and M. Houzet for helpful discussions.
D. Feinberg 
participated in the early BTK model calculations for
the NISIN structure. We thank also B. Dou\c{c}ot and J. Ranninger
for discussions on collective modes, and acknowledge
useful comments during a blackboard informal seminar at Grenoble.
The Centre de Recherches sur les Tr\`es Basses Temp\'eratures
is associated with the Universit\'e Joseph Fourier.

\end{document}